\documentclass{aa}
\usepackage{newtxtext,newtxmath}
\usepackage[colorlinks,citecolor=blue]{hyperref}

\usepackage[T1]{fontenc}
\usepackage{ae,aecompl}

\usepackage{graphicx}   
\usepackage{amsmath}    

\usepackage{bm}         

\begin{document}

\title{\textcolor{black}{Boxy/peanut} shaping of a mature galactic bar in action-angle space}

\author{
  Viktor D. Zozulia\inst{1,2}
\and 
  Anton A. Smirnov\inst{2}
\and 
  Natalia Ya. Sotnikova\inst{1,2}
\and 
  Alexander A. Marchuk\inst{1,2}
}
\institute{
  St. Petersburg State University,
  Universitetskij pr.~28, 198504 St. Petersburg, Stary Peterhof, Russia \\
  \email{n.sotnikova@spbu.ru}
  \and
  Central (Pulkovo) Astronomical Observatory of RAS, Pulkovskoye Chaussee 65/1, 196140 St. Petersburg, Russia
}


\date{Received XXX; accepted YYY}





\abstract
{
We study vertical resonant trapping and resonant heating of orbits. These two processes both lead to the growth of a boxy/peanut-shaped bulge in a typical $N$-body model. For the first time, we study this by means of the action variables and resonant angles of the actual orbits that compose the model itself. We used the resonant angle instead of the frequency ratio, which allowed us to clearly distinguish between these two processes in numerical simulations. We show that trapping and heating occur simultaneously, at least at the stage of a mature bar, that is, some orbits quickly pass through vertical resonance while at the same time, a substantial number of orbits remains trapped into this stage for a long time. Half of all bar orbits spend more than 2.5 Gyr in vertical resonance over an interval of 4 Gyr. Half of the orbits trapped into the bar over the last 3 Gyr of simulation remain captured in vertical resonance for more than 2 Gyr. We conclude that in the later stages of the bar evolution, the process of vertical trapping dominates in the ongoing process that causes the boxy/peanut shape of a bar in a typical $N$-body model. This contradicts the results of several recent works.
}


\maketitle



\section{Introduction}
\label{sec:intro}
When they are viewed edge-on, the central regions of many disk galaxies exhibit a prominent structure that bulges out from the midplane of the disk and resembles either a box or, in some cases, a peanut (e.g., NGC 128, NGC 4013). These structures, usually called boxy/peanut-shaped (B/PS) bulges, are directly observed in 30\%\,--\,40\% of the edge-on disk galaxies in the local Universe~\citep{Lutticke_etal2000,Erwin_Debattista2013,Yoshino_Yamauchi2015,Erwin_Debattista2017,Kruk_etal2019,Marchuk2022}, and some candidates have been identified up to redshift $z\sim1$~\citep{Kruk_etal2019}. 
Early numerical~\citep{Combes_Sanders1981,Combes_etal1990,Pfenniger_Friedli1991,Raha_etal1991} and kinematic~\citep{Kuijken_Merrifield1995, Bureau_Freeman1999, Merrifield_Kuijken1999, Veilleux_etal1999,Chung_Bureau2004} studies showed that B/PS bulges originate from the bars and are vertically extended parts of the bars. \cite{Combes_Sanders1981} were the first to show that a flat bar, shortly after its appearance, begins to grow in the vertical direction and acquires the B/P shape. Extensive orbital studies of numerical models carried out later \citep{Skokos_etal2002a,Patsis_Katsanikas2014a,Portail_etal2015b,Abbott_etal2017,Parul_etal2020} proved that B/PS bulges are built by the same orbits that populate the bars, that is, B/PS bulges share their building blocks with bars\footnote{Although not all bar orbits become B/PS bulge orbits.}. 
\par
Although the connection between bars and B/PS bulges is well known, the debate is ongoing as to why initially flat orbits rise above the disk plane and what causes this~\citep{Sellwood_Gerhard2020, Li_etal2023}. There are two different approaches. 
The first approach is based on the commensurability of three orbital frequencies (azimuthal $\Omega$, radial $\kappa$, and vertical $\omega_z$), which leads to resonant interaction. As a result, the orbits experience an average nonzero lifting force~\citep{Combes_etal1990,Quillen2002, Quillen_etal2014}.
The second concept considers the emergence of instability due to the collective response of the orbits to a vertical disturbance~\citep{Raha_etal1991}. This instability is often called buckling instability in the case of a growing bar and is similar to the firehorse instability. 
In the case of resonant interactions, the important resonance is the vertical inner Lindblad resonance (vILR). In this region of the disk, $\omega_z\approx2(\Omega-\Omega_\mathrm{p})$, and $\Omega_\mathrm{p}$ is the bar pattern speed. \citet{Combes_etal1990} noted that the location of this particular resonance almost coincides with the location of another resonance, namely the inner Lindblad resonance (ILR, $\kappa\approx2(\Omega-\Omega_\mathrm{p})$), which is known to play an important role in the formation and growth of the bar. Because the locations of resonances coincide, the elongated orbits of a bar, which are initially almost completely confined to the disk plane, sooner or later rise above the disk plane, and the orbits themselves become part of the B/PS bulge. 
\par A simple Hamiltonian model describing orbital trapping by vILR was first presented by~\cite{Quillen2002} and was based on an analogous model developed for in-plane resonance by~\cite{Contopoulos1975}. 
The ideas of~\cite{Quillen2002} were later expanded based on the fact that the secular evolution of a bar, driven by interaction with the disk and halo particles at the corotation resonance (CR; $\Omega=\Omega_\mathrm{p}$) leads to an increase in the bar size \citep{Athanassoula2003}. 
This suggests that the resonance moves outward and elevates particles from the midplane along the path~\citep{Quillen_etal2014}. 
As the resonance moves farther from the centre, increasingly more orbits are lifted from the disk plane and eventually form an impressive B/PS bulge. This process can briefly be described as resonant heating or a resonant passage, when we consider the orbits in the bar reference frame.
\par
\par
\citet{Quillen_etal2014} showed how resonant heating works using limited model examples. \citet{Sellwood_Gerhard2020} studied several self-consistent $N$-body models and one model with an artificially imposed vertical symmetry. The authors followed the secular evolution of the orbital frequencies of a random sample of bar particles in each model and concluded that although peanut shapes formed in all models, each model individually exhibited the effect of buckling, resonant heating, and resonant trapping.
\par
Although three distinct mechanisms have clearly been demonstrated that each contribute to the growth of the B/PS bulge, several points regarding them are still unclear. It is unclear whether each of these mechanisms operates only when certain conditions are met, and if not, whether they can operate simultaneously or if each is associated with a specific stage in the bar life. It also remains to be determined which of the two is more important for the B/PS bulge growth. It is critical to answer these questions if we wish to understand how B/PS bulges grow in real galaxies.
\par 
In this work, we demonstrate that at least the long-term resonant trapping and resonant passage (heating) that were studied in particular by \citet{Sellwood_Gerhard2020} occur simultaneously in a typical $N$-body model. In addition, resonant trapping can lock the orbits for very long periods of time, up to several billion years, while the bar itself continues to evolve slowly. To demonstrate this, we analyzed the evolution of a sample of orbits that were trapped into the bar in such a model in terms of both actions and resonant angles. We used the techniques developed in our previous work~\citep{Zozulia_etal2024}. Although the resonant angles arise naturally in theoretical studies, it is an entirely new approach to employ them to describe the evolution of orbits in $N$-body simulations. This allows us to directly identify orbits that are captured or escape the resonance. As we show below, the commensurability of the frequencies at some point in time is clearly insufficient to state that the orbit is in resonance, and it is therefore important to study the corresponding angles. \textcolor{black}{We would also like to note that by their theoretical definition, the actions~\citep{Lichtenberg_Lieberman_1992} should be constants of motion. However, when a system is perturbed by a strong bar, the actions defined for an axisymmetric potential are no longer conserved and are not really proper actions in the original sense. This arises because in this study and in some recent studies that considered the influence of the bar ~\citep{Binney_2018, Binney_2020, Trick_2021, Trick_2022}, the actions were calculated using the polar coordinates of particular orbits. These actions are sometimes called unperturbed actions~\citep{Binney_2018, Binney_2020} because they correspond to actions of the angularly averaged system (i.e., an axisymmetric system). To investigate the behavior of unperturbed actions on different timescales, we time-averaged these actions over different time intervals (Sec.~\ref{sec:measuring_actions}).}


 Our model contains an early buckling episode. We excluded this episode and its role in the uplift of orbits in the central regions in the early stages of bar evolution here and focused on the later stages, when buckling no longer affects the orbits that join the bar.
\par

\section{Simulations and action variables}
\label{sec:sim}

\subsection{Numerical model}
\label{sec:nbody}

To study how the orbits acquire their vertical extensions, we used the same numerical model as in~\cite{Zozulia_etal2024}, where we studied the process of orbital trapping in terms of the actions and frequencies of the orbits. Here, we briefly describe the properties of the model and refer to the cited work for more details.
\par 
The model initially consisted of two physical components: a pure stellar disk (exponential) and a spherical dark halo with an Navarro–Frenk–White density profile~(NFW, \citealt{NFW}). A numerical implementation of the model was prepared using the subroutine \texttt{mkgalaxy}~\citep{McMillan_Dehnen2007}. The components were then evolved in a self-consistent manner for about 8~Gyr using the~\texttt{gyrfalcON} code~\citep{Dehnen2002}, and a strong and fast bar emerged at the center of the disk at about 1~Gyr. Almost immediately after the formation of a bar, its inner part become vertically extended, that is, a B/PS bulge emerged. After the bar\footnote{In the further analysis, we identify the bar as a structure assembled from so-called abnormal orbits, that is, orbits that change angular momentum synchronously with the rate of orbital precession \citep{Zozulia_etal2024}.} formation, both the flat bar and the B/PS bulge continued to increase in size, capturing new disk orbits and subsequently lifting them out of the disk midplane~\citep{Zozulia_etal2024}. 
\par
Throughout this work, we use the natural system of units of the $N$-body model, in which the gravitational constant $G=1$, the disk mass $M_\mathrm{d}=1$, and the initial exponential scale of the disk $R_\mathrm{d}=1$. \textcolor{black}{One time unit corresponds to 13.8 Myr when the disk total mass is $M_\mathrm{d}=5\times 10^{10}M_\mathrm{\sun}$ and the scale length $R_\mathrm{d}=3.5$ kpc.}

\subsection{Frequencies and action-angle variables}
\label{sec:measuring_actions}

Following our previous work \citep{Zozulia_etal2024}, we estimated three \textcolor{black}{unperturbed (see \citealp{Binney_2018,Binney_2020})} actions $\bm{J}=(J_\mathrm{R},\,J_z,\,L_z)$ and frequencies $\bm{\Omega} = (\kappa,\,\omega_z,\,\Omega)$ on different timescales, namely medium-term and secular values. Medium-term values were calculated either on the scale of one radial oscillation ($J_\mathrm{R},\,\kappa$, and $L_z,\,\Omega$) or of one oscillation along the $z$-axis ($J_z,\, \omega_z$). On the same timescale, we found the angles $(\theta_R,\,\theta_z,\,\theta_\phi)$ by integrating the frequencies over time.  The secular values ($\bm{J}$ and $\bm{\Omega}$) were obtained on the timescale of one oscillation of the medium-term value. Almost all quantities in this work are secular unless otherwise noted.\par
We explored the relation between the vertical bar structure and the vILR. To do this, we defined four different behaviors of the resonant angle $\theta_z - \theta_R$:
\begin{itemize}
    \item[1, 2] $\dot{\theta}_z > \dot{\theta}_\mathrm{R}$ or $\dot{\theta}_z < \dot{\theta}_\mathrm{R}$. Circulation of the resonance angle with a positive or negative angular speed. 
    \item[3.] "In resonance". This designation is used for orbits that librate near the resonance angles $0$ or $\pi$ (for banana orbits) and $\pi/2$ or $3\pi/2$ (for anti-banana orbits).
    \item[4.] "Passage". This notation is used for orbits that are in the process of changing their behavior (changing the direction of circulation) and are also near the resonance (but do not become stuck in it).
\end{itemize}
In Sect.~\ref{sec:orbital_types} we give specific examples of the orbits for each of the cases described above. Using our classification of orbital behavior, we not only determine whether the orbit is in resonance or in the passage, but can also estimate the time it spends in one or the other state. 
\par
A detailed description of the calculations of the action-angle variables and the determination of the different behavior of the resonant angle is presented in Appendix \ref{sec:appendix}.

\par

\section{Evolution of the vertical bar structure in action-frequency space}
\label{sec:main}

\subsection{General structure of a bar in action-frequency space}
\label{sec:bar}

\begin{figure*}
\centering
\includegraphics[width=0.95\linewidth]{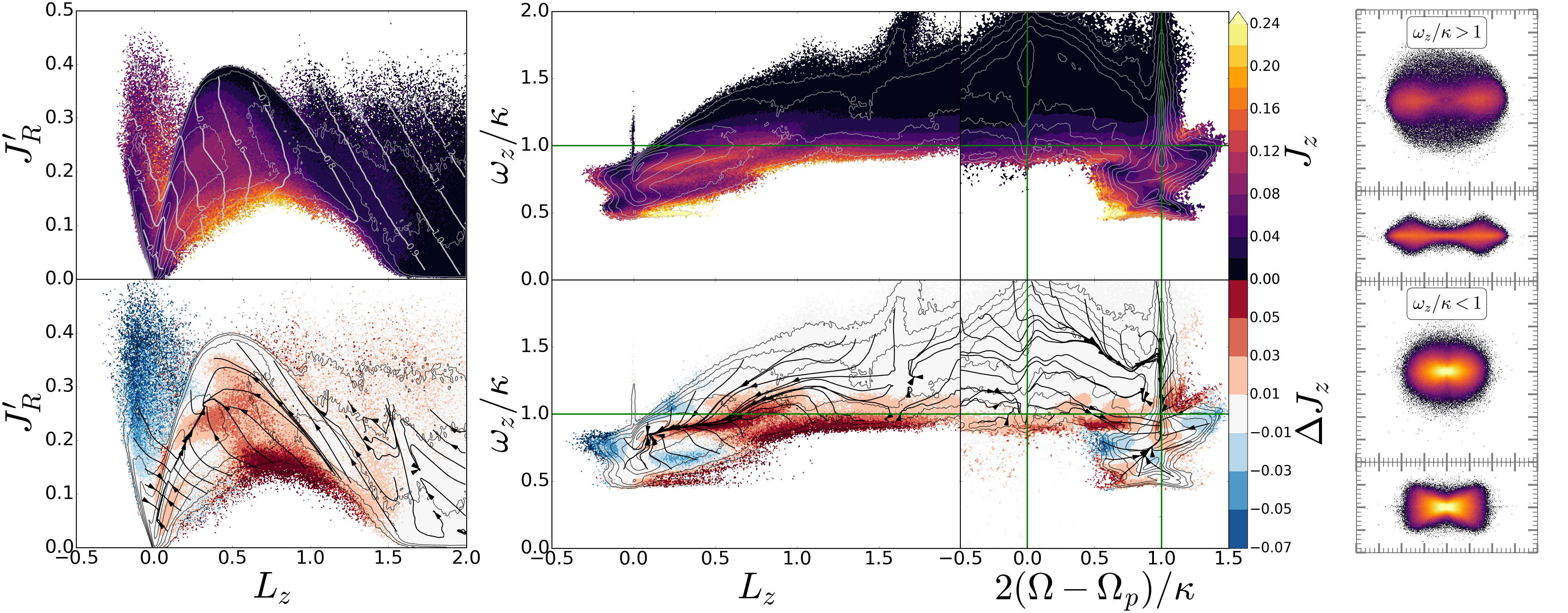}
\caption{Vertical action and its evolution. \textit{\textcolor{black}{Left and middle columns:}} 2D maps of the vertical action $J_z$ plotted in different coordinates, ($L_z$, $J_\mathrm{R}'=J_\mathrm{R}-\theta(-L_z)L_z$) (\textit{first column}),  ($L_z$, $\omega_z/\kappa$) (\textit{second column}), and ($2(\Omega-\Omega_p)/\kappa$, $\omega_z/\kappa$) (\textit{third column}). \textit{Top row:} Average value of the action in the pixel. \textit{Bottom row:} Mean changes in $J_z$ from $t=300$ to $t=400$ (actions and frequencies refer to the moment $t=400$). The vector fields in the \textit{bottom rows} demonstrate the average flow of orbits in given planes over five time units (from $400$ to $405$). The thin gray lines in each plot represent isodensity contours. The thick gray lines in the \textit{top left plot} correspond to the lines with the same value of the adiabatic invariant $J_v=J_\mathrm{R}'+J_z+L_z/2$. \textcolor{black}{The vertical green lines in the \textit{middle column} indicate the location of the CR and ILR.} \textit{Right column:} Snapshots ($xy$ and $xz$) of the bar orbits (abnormal orbits; see \citealt{Zozulia_etal2024}) at $t=400$ with different frequency ratios $\omega_z/\kappa>1$ (\textit{top} two plots) and $\omega_z/\kappa<1$ (\textit{bottom} two plots).} 
\label{fig:allbar_maps}
\end{figure*}
In this section, we explore the structure of our model in action--frequency space
at the stage of a mature bar ($t=400$) that gradually grows in both directions, in-plane and vertical. We specifically focus on the vertical action $J_z$, which is tied to the vertical excursions of the orbits, and which changes when the orbits become trapped or heated by the resonance. We also note that by the time moment $t=400$, the episode of early buckling ($t\approx 180$) has long passed and does not affect the evolution of the vertical bar structure in the period under study.
\par
The first three top plots of Fig.~\ref{fig:allbar_maps} demonstrate color-coded values of the vertical action $J_z$ in the maps ($L_z,\,J'_\mathrm{R}$), ($L_z,\,\omega_z/\kappa$), and ($2(\Omega-\Omega_\mathrm{p})/\kappa,\,\omega_z/\kappa$), from left to right, respectively.
In the three bottom plots, on similar maps as in the top plots, color gradations show the changes $\Delta J_z$ over the time period from $t=300$ to $t=400$. Areas with an average increase in $J_z$ are marked in red, while areas in which $J_z$ decreases are shown in blue. 
The displacements (flows) of the orbits are presented in the form of a vector field in the same maps. These displacements were calculated based on the changes in the density of the orbits in a pixel over a period of time from $t=400$ to $t=405$, and they indicate where the ensemble of orbits tends to move in phase space over short periods of time.
\par
In the top left plot in Fig.~\ref{fig:allbar_maps}, most bar orbits are located below an outline resembling an inverted parabola and have $0\la L_z\la 1.5$. On the right, the parabola is limited by the envelope that corresponds to the constant value of the adiabatic invariant $J_v=0.8$ (one of the gray lines). Flat orbits with low values of $J_z$ are located here (dark blue and black in the color bar). These orbits have the highest values of the adiabatic invariant $J_v$ of the orbits that only recently became trapped into the bar\footnote{This is evident from the results of Sect.~\ref{sec:trapped}.}. Orbits that joined the bar earlier (with lower values of $J_v$) have already increased $J_z$. In the bottom left plot, we show in the same coordinates that the orbits move toward lower values of $L_z$ and higher values of $J'_\mathrm{R}$ in general, and their $J_z$ increases.
\par
The encounter of the orbit with the vILR leads to an increase in $J_z$. 
In terms of frequencies, the vertical resonance condition\footnote{Since the orbits are already in the bar, their frequencies satisfy the condition $2(\Omega-\Omega_\mathrm{p})=\kappa$, and the condition $\omega_z/\kappa=1$ corresponds to $2( \Omega-\Omega_\mathrm{ p})=\omega_z$.} for the bar orbits can be written as $\omega_z/\kappa=1$. The second upper plot in Fig.~\ref{fig:allbar_maps} shows that the orbits in the region $0\la L_z\la 1.5$ with high values of $J_z$\footnote{These orbits belong to the bar.} (purple and orange) lie in the map below the horizontal line $\omega_z/\kappa=1$, that is, they have already passed through the vILR. The orbital flows are generally directed below the horizontal line $\omega_z/\kappa=1$ (bottom plot in the same phase space). 
\par
In the third top map in Fig.~\ref{fig:allbar_maps}, most of the orbits in the bar are located along the vertical stripe near the line $2(\Omega-\Omega_\mathrm{p})/\kappa=1$. Orbits with the highest $J_z$ have $\omega_z/\kappa<1$. For these, the encounter with the vILR has already taken place. Flat orbits (low values of $J_z$) trapped into a bar have $\omega_z/\kappa>1$. Some of the orbits have $\omega_z/\kappa=1$ and appear to be in both in-plane and vertical resonances simultaneously. However, based on this fact alone, we cannot conclude whether they are stuck in the resonance for a long time or are in the process of passage. The most interesting changes are visible in the vector map ($2(\Omega-\Omega_\mathrm{p})/\kappa,\,\omega_z/\kappa$). Here, the flows are generally directed from left to right toward the line $2(\Omega-\Omega_\mathrm{p})/\kappa=1$, and then the arrows indicate a descent to the vILR ($\omega_z/\kappa=1$) and immediately below it, where the region with the greatest change in $J_z$ is located.
\par
The bar orbits that are in the process of passage through the vILR ($\omega_z/\kappa>1$), and those that have already passed through it ($\omega_z/\kappa<1$), are assembled into morphologically different structures (Fig.~\ref{fig:allbar_maps}, fourth column of the four snapshots). The latter orbits constitute a more compact structure in the in-plane projection and have a pronounced B/P shape in the vertical direction. Orbits with $\omega_z/\kappa>1$ join the bar later. They are more extended and have not yet had time to elevate significantly. Their B/P shape is weakly expressed.
\par
It should be noted that the maps of the $J_z$ distributions in a particular time moment in Fig.~\ref{fig:allbar_maps} are the result of the action of all possible mechanisms of orbital elevation over a long time interval. These mechanisms can operate on different timescales. In Sect.~\ref{sec:trapped} we analyze the fate of initially flat orbits (those with $J_z({t=300})<0.05$) that were only recently trapped into the bar to distinguish between the mechanisms of vertical elevation of orbits in momento.

\subsection{Trapped orbits in action-frequency space}
\label{sec:trapped}

\begin{figure*}
\centering
\includegraphics[width=0.88\linewidth]{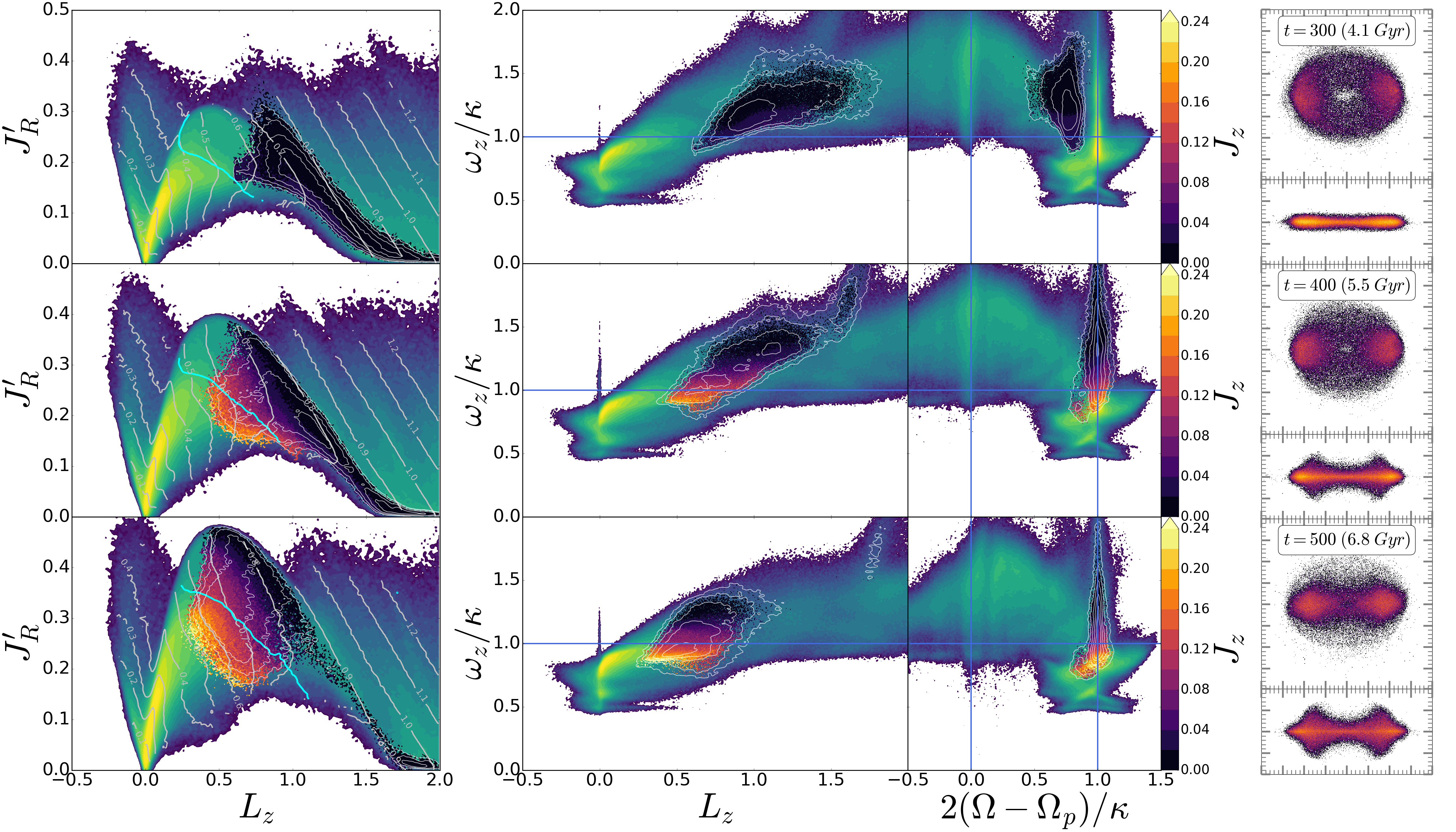}
\caption{Vertical action and its evolution for captured orbits. \textit{\textcolor{black}{Left and middle columns:}} 2D density maps (bar and disk) in the background and 2D maps (superimposed in the foreground) of $J_z$ for orbits that were flat ($J_z<0.05$) at $t=300$ and trapped into the bar over the time interval of $t=300-400$. \textcolor{black}{Different time moments are shown from top to bottom, $t=300$ (\textit{top row}), $400$ (\textit{middle row}), and $500$ (\textit{bottom row}).} The thin gray lines in each plot represent the isodensity contours of trapped orbits. In the \textit{left column}, the thick gray lines correspond to isolines of the same value of the adiabatic invariant $J_v=J_\mathrm{R}'+J_z+L_z/2$, and the thick cyan line corresponds to the frequency ratio isoline $\omega_z/\kappa=1$. \textcolor{black}{The vertical blue lines in the \textit{middle column} indicate the location of the CR and ILR.} \textit{Right column:} Snapshots ($xy$ and $xz$) of the trapped orbits at the corresponding time moments.}
\label{fig:trapped_orbits}
\end{figure*}

Figure~\ref{fig:trapped_orbits} (the first three maps in each row) shows the density of all orbits by yellow-green gradations in color in a pixel for the same phase coordinates as in Fig.~\ref{fig:allbar_maps}. We overlaid the location of the trapped orbits in each map, and the black-orange-yellow color gradations (as in Fig.~\ref{fig:allbar_maps}) correspond to $J_z$ for these orbits. The maps are given for three time moments. $t=300$: The orbits in question have not yet fallen into the trap of the bar. On the right, they are limited by the isoline $J_z \approx 0.8$, and in the bar itself, there are no orbits yet with these values of $J_z$ (first plot on the left). They are also enclosed left of the vertical line
$2(\Omega-\Omega_\mathrm{p})/\kappa=1$ in a black region (flat orbits), outlined by light blue isodenses (third plot from the left). $t=400$: The orbits have already been captured by a bar. $t=500$: The orbits have been in the bar for a long time and continue to evolve.
\par
The changes in the phase maps (Fig.~\ref{fig:trapped_orbits}) correspond to the general trends noted in Sect.~\ref{sec:bar}. The orbits of the entire bar and the orbits trapped into the bar over the time interval $t=300-400$ both increase in $J'_\mathrm{R}$ (first column). The area overlaid on the maps, which only consists of trapped orbits, moves toward lower $L_z$ values (first and second columns). With increasing $J_z$, the trapped orbits move below the line $\omega_z/\kappa=1$ (in the first column of Fig.~\ref{fig:trapped_orbits}, this line is plotted in blue). The plots in the third column show that the orbits gradually propagate toward the in-plane resonance (from left to right) and then descend along a vertical line $2(\Omega-\Omega_\mathrm{p})/\kappa=1$, increase in $J_z$, with the highest values of $J_z$ achieved for $\omega/\kappa<1$ ($t=500$).
\par
The change in the morphology of the entire ensemble of trapped orbits is visible in the plots in the rightmost column of Fig.~\ref{fig:trapped_orbits}. The edge-on snapshot for $t=400$ closely resembles the edge-on snapshot in Fig.~\ref{fig:allbar_maps} for $\omega_z/\kappa>1$, which corresponds to orbits that mostly have not yet entered into vILR. Although these orbits have increased $J_z$, they do not protrude too much from the midplane.
\par
Since $\omega_z/\kappa=1$ is a condition for vILR, the orbits with $\omega_z/\kappa<1$ are those that have already passed through the vILR. 
At the same time, the blue isodenses at $t=500$ (Fig.~\ref{fig:trapped_orbits}, third column) show that pixels with the highest density for the entire ensemble of captured orbits are not located in the area with the highest $J_z$ values, but rather fall in the $\omega_z/\kappa=1$ region. 
This is the first indication that even at the stage of a mature bar, there may be many orbits among the orbits that newly joined the bar that are directly trapped in vILR. In the next section, we study orbital trapping in vILR in more detail based on the long-term behavior of the resonant angle rather than based on the frequencies. We classify the four main types of orbits with respect to the vILR, and we provide statistics on orbital types.

\subsection{Orbital classification with respect to the vILR}
\label{sec:orbital_types}

\begin{figure*}
\centering
\includegraphics[width=0.49\linewidth]{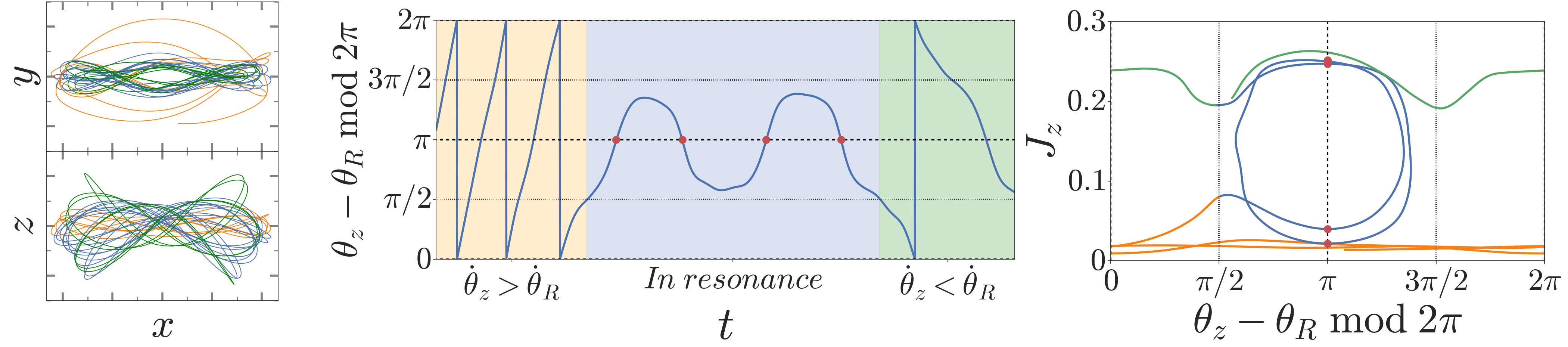}
\includegraphics[width=0.49\linewidth]{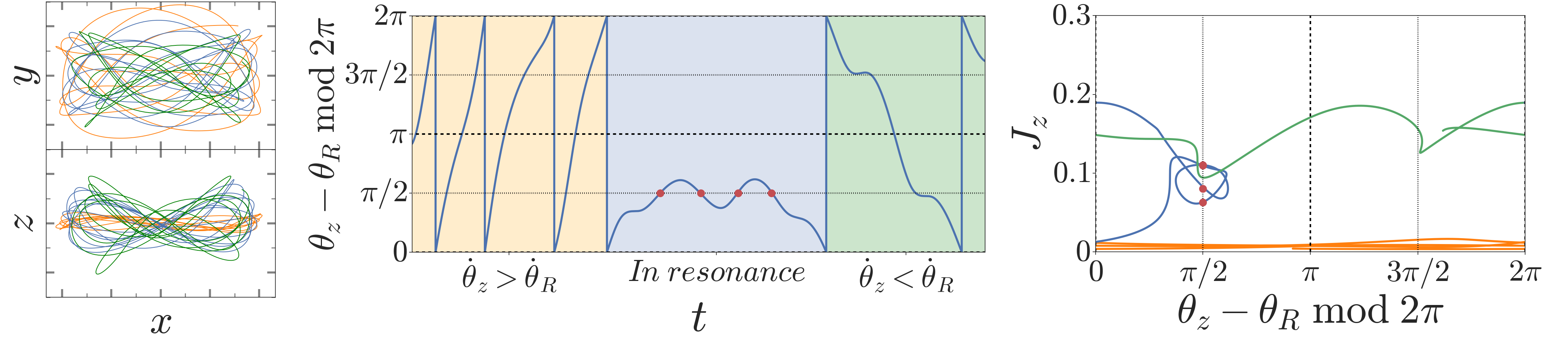}\\
\includegraphics[width=0.49\linewidth]{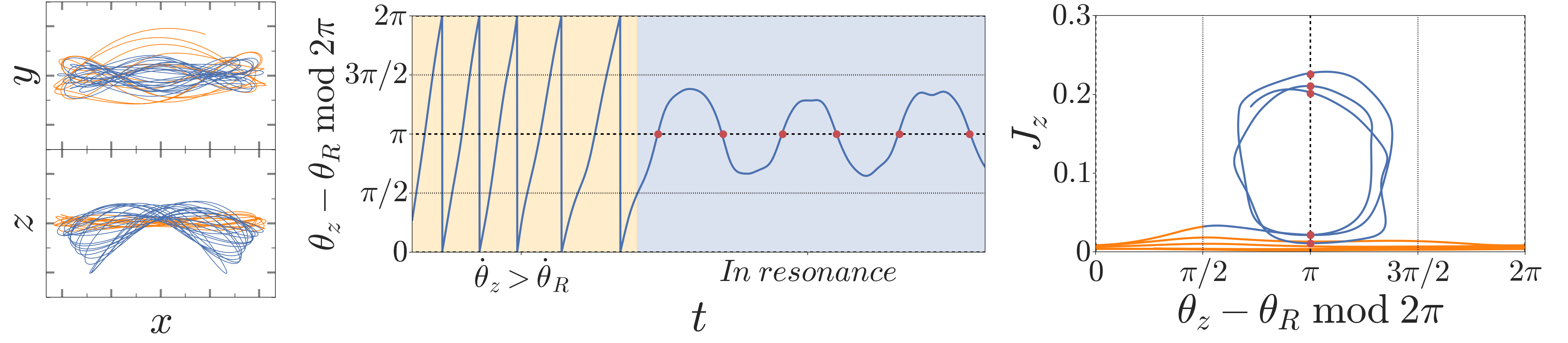}
\includegraphics[width=0.49\linewidth]{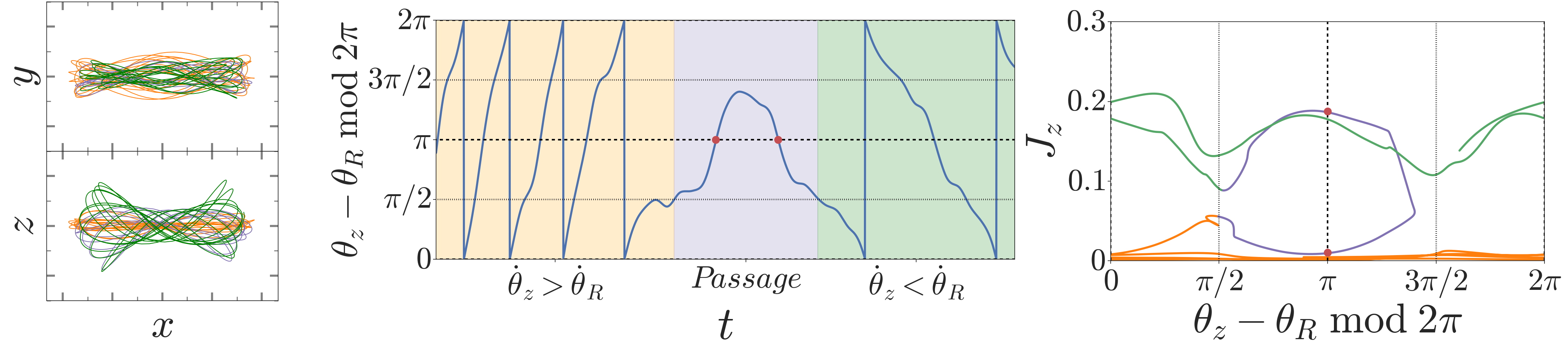}\\
\caption{Four examples of different orbital behaviors in our $N$-body simulations. For each orbit, \textit{from left to right}, we show snapshots ($xy$ and $xz$), the change in the resonant angle $\theta_z-\theta_\mathrm{R}$ with time, the orbit in the plane ($\theta_\mathrm{res},\,J_z$), where $J_z$ is the midterm vertical action. The \textit{top left and right} orbits remain a long time in vILR and librate around $\theta_z - \theta_R \sim \pi$ (banana orbit) and $\pi/2$ (anti-banana orbit), respectively, and then leave it. The \textit{bottom right} orbit is trapped into vILR and also librates around $\theta_z - \theta_R \sim \pi$ until the end of the integration time. The \textit{bottom left} orbit passes through the resonance around $\theta_z-\theta_\mathrm{R}=\pi$. \textit{Orange} and \textit{green} in each plot correspond to the time when $\theta_z-\theta_\mathrm{R}$ increases and decreases, respectively. \textit{Blue} and \textit{purple} correspond to the stage when the orbit is in resonance or passes through it, respectively.}
\label{fig:resonance_orbits}
\end{figure*}

In Sect.~\ref{sec:measuring_actions} we described how four different types of orbital behavior with respect to the vILR can be described. Considering the vILR and the resonant angle $\theta_{\mathrm{res}}=\theta_z-\theta_\mathrm{R}$, these four types correspond to 1) circulations with a monotonic increase in $\theta_\mathrm{res}$; 
2) the same, but with monotonically decreasing $\theta_{\mathrm{res}}$; 
3) libration around the resonant angles 
$\theta_\mathrm{res}=0$ or $\pi$ 
(banana orbits) and  $\pi/2$ or $3\pi/2$, 
(anti-banana orbits); and 4) the passage through the resonance ($\theta_{\mathrm{res}}$ changes sign in a particular manner). By studying the behavior of the resonant angle, it is possible  a) to determine the status of the orbit at any point in time by tracking the history of its resonant angle back and forth in time; b) to analyze how many orbits are in one state or another in relation to vILR at a given time; and c) to determine the time the orbit stays in resonance or in passage throughout the entire interval under consideration, and to give the statistics of this parameter.
\par
To illustrate how the process of trapping in vILR and passage through it occurs in a realistic $N$-body galactic model, we determined the mode of the resonant angle behavior in the time interval $t=250-550$, that is, after buckling. We did this for bar orbits and for recently trapped orbits.
\par
Figure~\ref{fig:resonance_orbits} gives specific examples illustrating the orbital behavior described above. For each orbit, we plot the projections on the $xy$ and $xz$ planes, the time evolution of the resonant angle $\theta_\mathrm{res}$ modulo $2\pi$, and the evolution of the orbit in the plane $(\theta_\mathrm{res},\,J_z)$\footnote{We use midterm values.}. The orbit, evolving, can change the modes of behavior of the resonant angle. Two upper and lower left sets of plots, each consisting of three panels, show three orbits that are trapped into vILR. As the galactic potential evolves, the orbits in the upper subplots escape from resonance and go into circulation mode (green), while their vertical action increases significantly. The orbit in the bottom left subplot shows a different behavior. It continues to librate around the resonance until the end of the simulation time, increasing and decreasing $J_z$ while maintaining the average value of $J_z$. In bottom right plot, we plot the orbit that rapidly passes through vILR without being locked into libration mode and leaves the disk midplane, increasing its vertical action $J_z$. 
\par
The orbits shown in Fig.~\ref{fig:resonance_orbits} are taken as an example. They are trapped into the bar at $t \approx 200-250$, that is, after the buckling episode. In the time interval $250-550$, the average value of the adiabatic invariant for them is close to 0.65-0.71. While they are flat, they are practically indistinguishable in morphology and extent along the $x$-axis, but their evolution in the vertical direction is different.
\par
Thus, in the $N$-body model that experienced buckling in an early stage of evolution, at a later stage we observe (at the level of individual orbits) the evolution of the vertical structure due to other mechanisms: trapping into resonance, long-term trapping and subsequent exit from resonance, and rapid passage through it.
\par
\begin{figure*}
\centering
\includegraphics[width=0.95\linewidth]{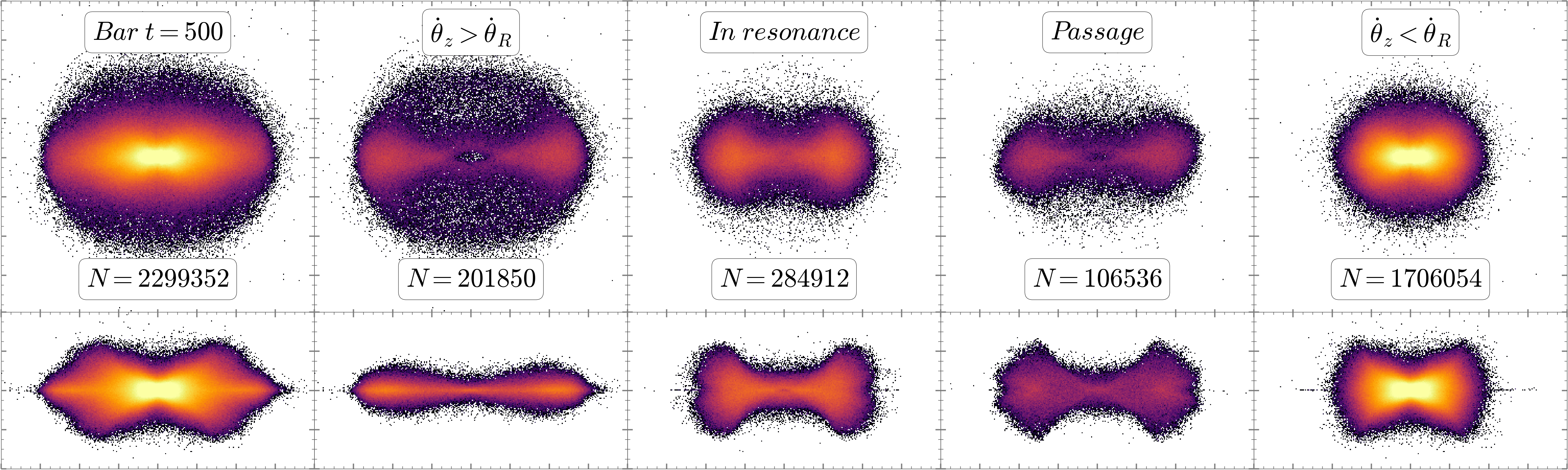}
\\
\includegraphics[width=0.95\linewidth]{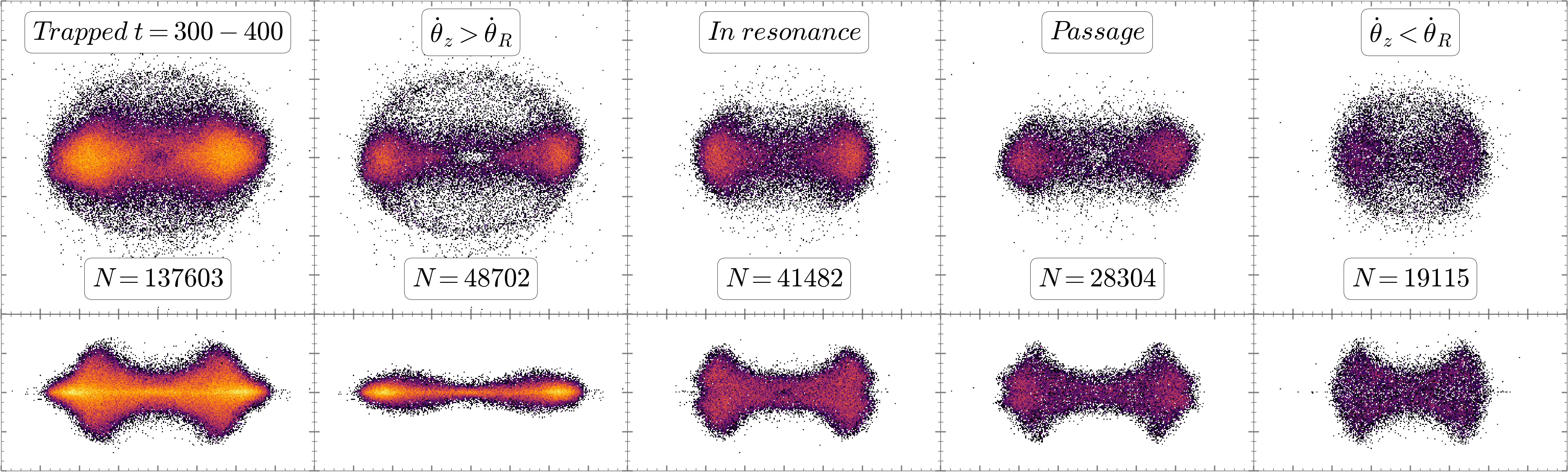}
\caption{Snapshots ($xy$ and $xz$) of the bar (\textit{top two rows}) and the structure assembled from the orbits trapped into the bar in the time interval $t=300-400$ (\textit{bottom two rows}) at $t=500$. \textit{The first column} shows all orbits under consideration. Structures assembled from orbits that increased and decreased their resonant angle $\theta_z-\theta_\mathrm{R}$ are displayed in \textit{the second and fifth columns}, respectively. \textit{The third and forth columns} show orbits in the vILR and passing through it, respectively.}
\label{fig:snap_bar_trapped}
\end{figure*}

Further, we considered the general structure assembled from the bar orbits and orbits trapped into the bar in $t=300-400$ (the same as in Sect.~\ref{sec:main}) on the $xy$ and $xz$ planes. We distinguished all orbits according to their behavior with respect to the vILR at time moment $t=500$. To determine the type of the corresponding orbital behavior at $t=500$, we do not need to know the entire history of the orbital evolution in the phase plane in the interval $t=250-550$. We track the orbit behavior over a certain period of time backward in time and, if necessary, forward, up to $t=550$.
Figure~\ref{fig:snap_bar_trapped} shows snapshots for each orbit type identified at time $t=500$. The top two rows refer to all orbits in the bar, and the two lower rows only show recently trapped orbits.

\begin{enumerate}
\item 
Orbits that only increase their angle $\theta_\mathrm{res}$
are collected into a flat bar (Fig.~\ref{fig:snap_bar_trapped}, second column) consisting of classical x1 and elliptical orbits. This is especially noticeable for orbits that are recently trapped into the bar. There are quite a few orbits with increasing $\theta_\mathrm{res}$ in a mature bar ($9\%$), but they are most numerous among the recently trapped orbits ($35\%$) and among all orbits with increasing $\theta_\mathrm{res}$ (trapped orbits of this type make up almost a quarter). These orbits continue to actively replenish the previously formed flat part of the bar. They have not yet entered the vILR, but they may lift out of the disk plane and enter the vILR later.
\item Orbits in the vILR that are trapped into the bar at the time interval $t=300-400$ account for 15\% of all orbits of this type in a bar that continue to be in the vILR for more than one libration period (Fig.~\ref{fig:snap_bar_trapped}, third column) at time $t=500$. Trapped orbits that are stuck in the vILR continue to participate in building this part of the B/PS bulge, which has an X-shaped structure that extends along the radius. Following the shape of the banana orbits, the distribution itself resembles two intersecting bananas. Among the orbits trapped into a bar in the interval $t=300-400$, this is the second most numerous type of orbit (30\%), although among all orbits in the bar, this type accounts for 12\% of the cases.
\item Orbits in passage (Fig.~\ref{fig:snap_bar_trapped}, fourth column) generally repeat the distribution of the resonant orbits. Near the moment $t=500$, they experience an episode of rapid passage through the vILR without turning around on the plane $(\theta_\mathrm{res},\,J_z)$, but with a sharp increase in $J_z$. The bar and the ensemble of trapped orbits contain about $5\%$ and $21\%$ of these orbits, respectively.
\item 
Orbits with a decreasing resonant angle (Fig.~\ref{fig:snap_bar_trapped}, fifth column), that is, those that are already out of the vILR, are the most numerous type of orbits included in the B/PS bulge ($74\%$), and in its thickest part. 
In the past, they may have been in resonance for some time or may have quickyl passed through the vILR.
We do not exclude the possibility that for some orbits of this type, the increase in $J_z$ was associated with an episode of buckling at $t\approx180$. However, this does not apply to orbits trapped into the bar in $t=300-400$. The number of recently trapped orbits among orbits of this type is relatively small ($14\%$), but they contribute precisely to the thickest and most compact part of the B/PS bulge. The structure into which they assemble is morphologically similar to the structure assembled from all orbits of this type.
\end{enumerate}

While the orbit is librating, $J_z$ oscillates around the mean value. Leaving the vILR, the orbit enters the circulation mode with the maximum value of $J_z$ (Fig.~\ref{fig:resonance_orbits}).
Thus, the thickest part of the B/PS bulge grows due to orbits that have passed through the vILR and begin to circulate, decreasing the resonant angle. 
However, orbits that have recently been trapped into the mature bar and entered circulation mode with a decreasing resonant angle are few, and they do not effectively replenish the thick parts of the bar. Among all the orbits of a bar of this type, they make up 1\%. In $100-200$ time units, about $2\cdot 10^{4}$ of $14\cdot 10^{4}$ orbits were added to the B/PS bulge in this manner. About half of the newly trapped orbits are still in or pass trough resonance (about $7\cdot 10^{4}$ of $14\cdot 10^{4}$). From the given statistics, it follows that the inner thickest part of the bar was formed long ago, in contrast to the thinner extended parts, which are effectively replenished by resonant orbits.

\subsection{Evolution of the orbital types}
\label{sec:time_in_resonance}
\begin{figure*}
\centering
\includegraphics[width=1\linewidth]{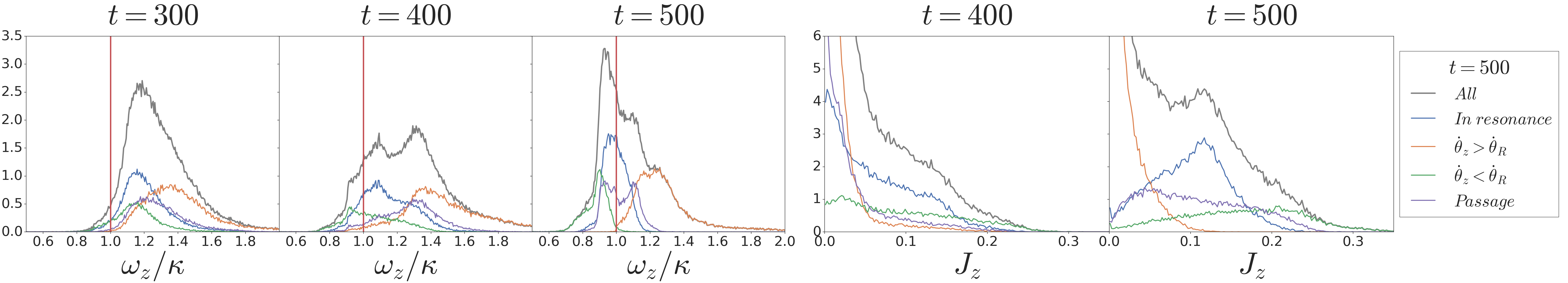}
\caption{1D distributions of the orbits trapped into the bar over the time interval $t=300-400$, which were flat ($J_z<0.05$) at $t=300$ (the same orbits as in Fig.~\ref{fig:trapped_orbits}): \textit{The three left plots} display the secular frequency ratio $\omega_z /\kappa$ at $t=300,\,400, and \,500$, respectively. \textit{ The two right plots} correspond to the secular vertical action $J_z$ at $t=400$ and $t=500$. The gray lines show the distributions of all trapped orbits, and the other colors indicate orbits with different behaviors (the same as in Fig.~\ref{fig:resonance_orbits}) at $t=500$.}
\label{fig:hist_trappead_resonance}
\end{figure*}

In Sect.~\ref{sec:orbital_types} all orbits trapped into the bar from $t=300$ to $t=400$ were classified into four types with respect to the vILR at $t=500$. 
We present the change in $\omega_z/\kappa$ and $J_z$ for these types from $t=300$ to $t=500$.
Figure~\ref{fig:hist_trappead_resonance} shows the 1D distribution over these quantities\footnote{Below, we again use secular values of all quantities, in contrast to the values in Fig.~\ref{fig:resonance_orbits}.}. 
All types of orbits on average reduce the ratio $\omega_z/\kappa$ and increase $J_z$.
\par
Orbits that increase their resonant angle indeed have $\omega_z/\kappa>1$ and are flat in all time moments (orange lines in Fig.~\ref{fig:hist_trappead_resonance}). For these, $J_z<0.1$ even at $t=500$ with an explicit peak at $J_z=0$. In turn, orbits that decrease their angle at $t=500$ have a frequency ratio $\omega_z/\kappa<1$ and the widest distribution over $J_z$, spanning the range up to $J_z \approx 0.3$ and beyond (green lines in Fig.~\ref{fig:hist_trappead_resonance}).
\par
Orbits in resonance and passage through it (in Fig.~\ref{fig:hist_trappead_resonance}, blue and violet lines, respectively) are distributed around $\omega_z/\kappa \approx 1$ at $t=500$, but because these orbits librate, the distributions are smeared, and both distributions are poorly distinguishable in $\omega_z/\kappa$. For this reason, almost resonant orbits are more accurately identified by the evolution of their resonant angle and not by their frequency ratio\footnote{It should be noted that the distributions over $\omega_z/\kappa$ for the two extreme types of orbits, with increasing and decreasing resonant angles, do not intersect at $t=500$.}. Both orbit types have a wide range of secular $J_z$ from $0$ to $0.25$, with an average value of approximately $0.15$.
\par
\par
\begin{figure*}
\centering
\includegraphics[width=0.49\linewidth]{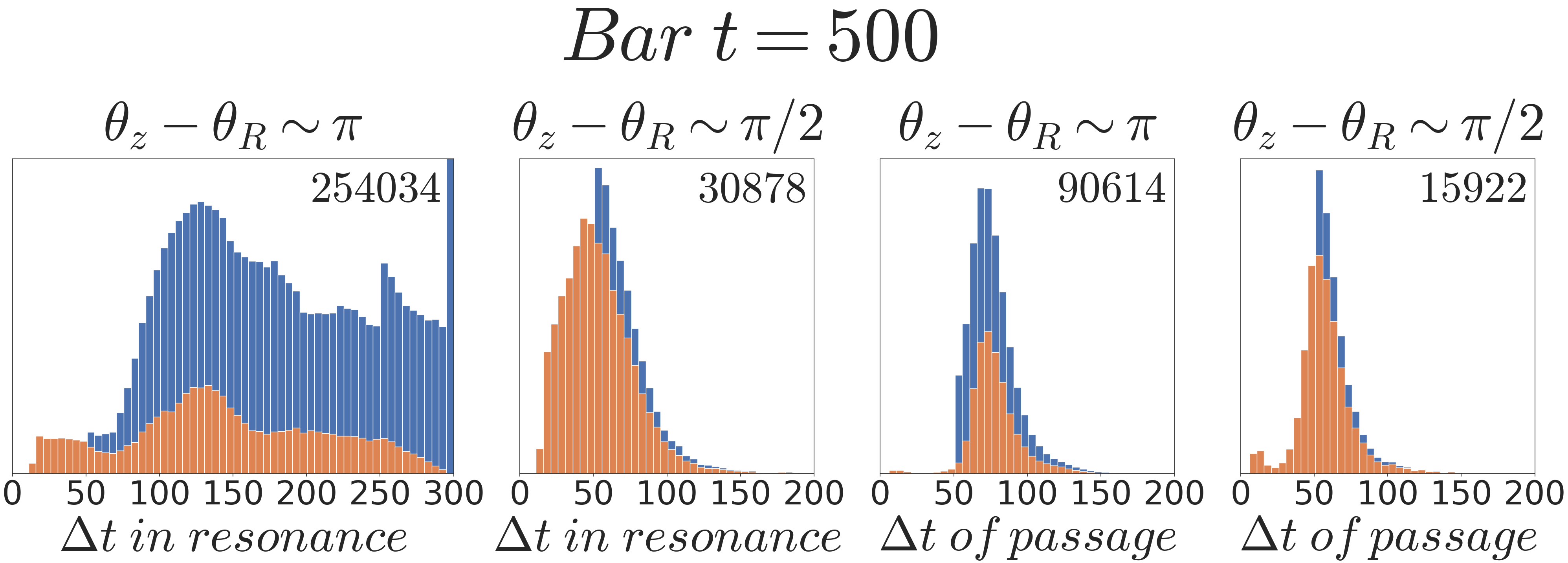}
\includegraphics[width=0.49\linewidth]{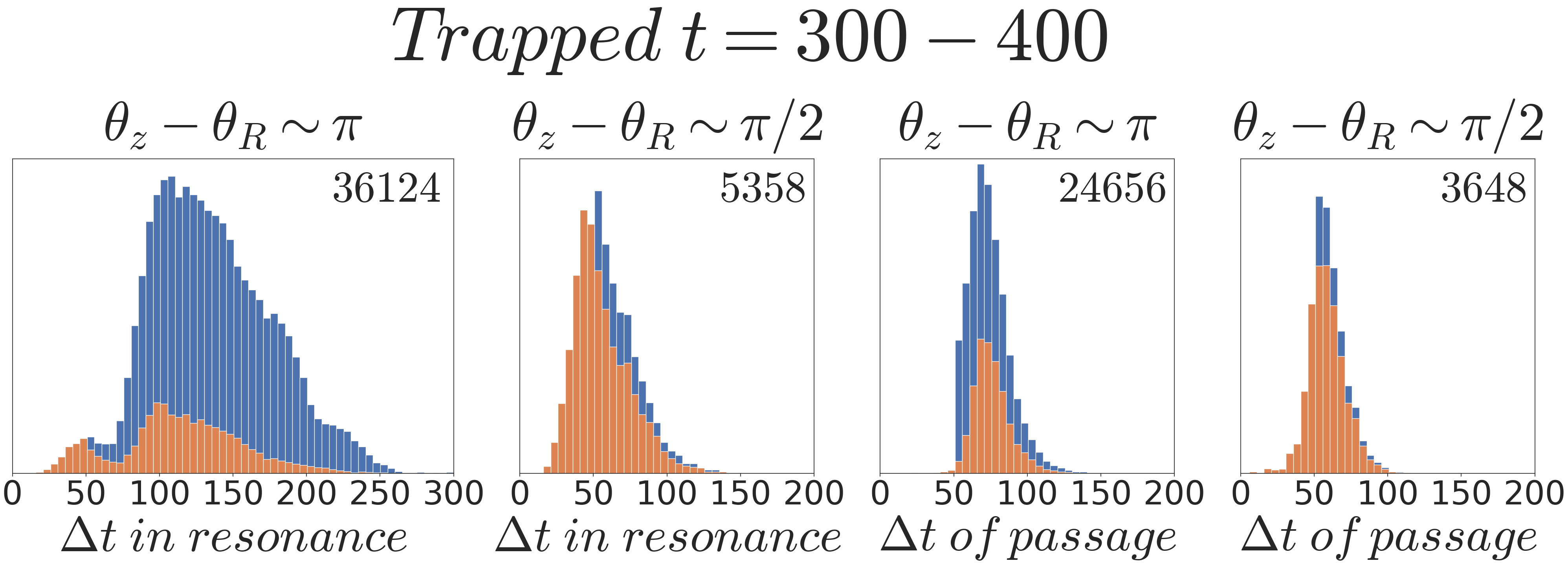}
\caption{Distributions of the time that the bar (\textit{four left plots}) and trapped (\textit{four right plots}) orbits spend in resonance or pass through it, around the angles $0,\,\pi$, and $\pi/2,\,3\pi/2$. Orbits with resonant angles of $0,\,\pi$ are combined and are designated in the plots as $\theta_z-\theta_\mathrm{R} \sim \pi$. The same holds for the angles $\pi/2,\,3\pi/2$. They are denoted as $\theta_z-\theta_\mathrm{R} \sim \pi/2$. The number of each type of orbit is indicated in the top right corner of each plot. The resonant and passing orbits are identified at $t=500$, while the time interval $t=250-550$ is being considered. The \textit{orange} distributions correspond to orbits that enter and leave resonance or pass through it during this time period. The \textit{blue} part of distributions corresponds to orbits that at the time $t=250$ were already in resonance and/or were still in resonance at moment $t=550$, and thus, they show a lower estimate for the time spent near the vILR.}
\label{fig:hist_t_in_resonance}
\end{figure*}

We considered the entire history of the orbital evolution and determined how long orbits stay near resonance in the time interval from $t=250$ to $t=550$ (the stage of a mature bar; after buckling).
The distributions over this time ($\Delta t$) are shown in Fig.~\ref{fig:hist_t_in_resonance} separately for all orbits in the bar and for orbits trapped in the time interval between $t=300$ and $t=400$. The classification of orbits by type (passage through resonance and libration near resonance) is made at $t=500$.
We also distinguish 
between banana orbits ($\theta_z-\theta_\mathrm{R} \sim \pi$ in Fig.~\ref{fig:hist_t_in_resonance}) and anti-banana orbits ($\theta_z-\theta_\mathrm{R} \sim \pi/2$ in Fig.~\ref{fig:hist_t_in_resonance}).
Anti-banana orbits are unstable \citep{Pfenniger_Friedli1991}, and we expect $\Delta t$ for them to be shorter than for banana orbits. They also leave resonance and enter the circulation mode earlier. 
\par
The orange part of distributions corresponds to the orbits that enter the resonant area (vILR) after $t=250$ and leave it after $t=500$, but before $t=550$, that is, we exactly know the time of entry and exit from resonance. These orbits cannot spend more than 300 time units of time near resonance.
\par
The blue part represents the remaining orbits. These orbits could be already in resonance at $t=250$ and/or were still in resonance at $t=550$. For orbits trapped into the bar in the time interval $300-400$, we can determine the moment of the vertical resonance encounter, but we cannot determine the moment of exit from vILR after $t=550$. 
It follows from the above that for all bar orbits and for orbits that were recently captured into the bar, the blue distributions show a lower estimate for the time spent near resonance because we do not analyze the model before $t=250$ for all bar orbits and after $t=550$ and cannot fix the moment of entry and/or exit from resonance. 
The sum of the orange and blue distributions gives all orbits in the bar or all trapped orbits ($t=300-400$). 
\par
The plots show that near the resonant angle $\sim \pi/2\,(3\pi/2)$, the orbits spend little time (about $50$ time units, or about 0.7 Gyr). In contrast, orbits that are in resonance near $\sim 0\,(\pi)$ are stable banana orbits and can be in a state of long-term libration around the resonance. Half of all bar orbits with a resonant angle $\sim 0\,(\pi)$ spend more than 180 time units in vertical resonance over a time interval from $t=250$ to $t=550$. Around $3\cdot 10^4$ orbits in the bar spend more than $300$ time units in resonance, or more than 4 Gyr, and some of them could have entered this mode from a buckling episode. However, in our simulations, `recently' trapped orbits cannot spend more than $200-250$ time units (2.8-3.5 Gyr) in resonance, because before time $t=300$, they were not in the bar and had a low $J_z$ value. Nevertheless, half of trapped orbits stay in the vILR for more than 135 time units (1.9 Gyr), that is, we observe them in the vILR most of the time.

\section{Discussion}
\label{sec:discussion}

An important preliminary step in describing resonance phenomena in a stellar disk is to use the frequency characteristics of the disk itself, which are determined by the model potential. However, the dynamic properties of individual orbits and their evolution in a changing potential can hardly be described within the framework of this approach. This is the general idea that we continue to develop in this work.
\par
\citet{Ceverino_Klypin2007} were perhaps the first to clearly demonstrate with examples of completely self-consistent models without imposed restrictions in the form of a frozen potential that resonances in the stellar disk (ILR, CR, and OLR) are not special regions in the disk, but collections of orbits with certain frequency characteristics. \citet{Parul_etal2020} used a similar approach to show that the brightest part of BP/S bulges, the so-called X-structures, is built from orbits with different spectral characteristics that are determined directly in a self-consistent $N$-body potential.
\par
Having supplemented frequencies with action variables, \citet{Zozulia_etal2024} introduced the concept of orbital mode, normal and abnormal, by analogy with Lynden-Bell's (\citeyear{Lynden-Bell1979}) normal and abnormal regions. \citet{Zozulia_etal2024} concluded that an orbit is trapped into a bar when the orbital mode changes from normal to abnormal, that is, when the angular momentum of the orbit and the rate of orbital precession begin to change synchronously, and not simply when the particle crosses the line of zero Lynden-Bell (\citeyear{Lynden-Bell1979}) derivative.
\par
Finally, we included into the analysis of self-consistent $N$-body simulations the concept of a resonant angle in addition to actions. The behavior of the resonant angle enables us to describe the vILR more accurately than via the commensurability of frequencies. We showed that the orbit that becomes stuck in or passes through the vILR is also a characteristic of the orbit itself, and not a consequence of its falling into a special region in the disk. Even at late stages of the bar evolution, the orbit can enter a long-term libration regime in the ($\theta_\mathrm{res},\,J_z$) plane. A significant fraction of orbits retain this behavior from the early stages of bar formation. Our findings are consistent with the results by \citet{Parul_etal2020}. In the models considered there, banana-shaped orbits (resonant orbits) occurred in a wide range of the Jacobi integral values, that is, both in the inner areas of the bar (formed earlier) and in the outer areas (joined to the bar later).
\par
When we qualitatively compare our results with the results of Hamiltonian analysis in \citet{Quillen2002,Quillen_etal2014}, individual particles indeed show the behavior described in both these works, that is, 2:1 orbits that librate around the resonance and orbits that pass through it. However, judging by the results of our study, the vILR moves and traps new orbits not along the disk, but in the space ($J_\mathrm{R},\,L_z,\,J_z$), as in Fig.~\ref{fig:trapped_orbits} (left column, blue line). 
\citet{Sellwood_Gerhard2020} gave an example of an orbit in resonance only for the case with artificially imposed vertical symmetry (their Fig. 13, upper plots). In its behavior, this orbit is analogous to our two upper orbits in Fig.~\ref{fig:resonance_orbits}. At the same time, we did not impose any restrictions on the model, but our orbits show a similar behavior at later stages, and most surprisingly, despite a long stay in resonance, the orbit can leave it (Fig.~\ref{fig:resonance_orbits}, upper left plots). The orbits in Fig.~\ref{fig:resonance_orbits}, including the orbits in passage, have similar values of the adiabatic invariant $J_v$ (0.6-0.7) and similar initial values of $J_z$, close to zero, but their fate is different. 


\section{Conclusions}
\label{sec:conclusions}

Considering the ensemble of orbits in the bar at the late stage of B/PS bulge building, we came to the following conclusions:

\begin{itemize}
    \item We identified the orbits that are being trapped into the bar at a particular time interval and tracked their evolution both in the plane and in the vertical direction. We studied the properties of corresponding orbits, including the action variables, the frequency ratio $\omega_z/\kappa$, and the resonant angle $\theta_\mathrm{res}=\theta_z-\theta_\mathrm{R}$. An analysis of the behavior of the resonant angle for all orbits in the $N$-body model on a long timescale was made for the first time. This allowed us to determine the various mechanisms whose action leads to the vertical bar shape in the stage of a mature bar, when the 
    bar experiences steady growth.
    \item We found that some orbits quickly pass through resonance (resonance heating) while others remain in the vILR for a long time. Surprisingly, the latter mechanism appears to be dominant in lifting the orbit from the disk midplane. Both types of interactions are observed for a significant number of orbits. Thus, we conclude that both mechanisms work simultaneously at least in our model. Since at the stage of a mature bar the episode of early buckling is long past, we conclude that orbit lifting arises only from resonant interaction. Because the properties of the orbits were calculated in a self-consistent model without imposed symmetry or in a frozen potential \citep{Sellwood_Gerhard2020}, we assume that both mechanisms work in real galaxies. 
    \item The orbits spend different times in resonance. The time of the resonance passage is very short for all orbit types (0.7 Gyr on average). Banana-like orbits can become stuck in the vILR for several billion years, however. A small group of orbits resides more than 4~Gyr in the vILR, but they do exist even though the properties of the bar change, that is, the bar slows down.
    \item Although our model includes a significant fraction of orbits in vILR in the late stages of bar evolution (Fig.~\ref{fig:snap_bar_trapped}), the thickest part of the bar is dominated by orbits in circulation mode with decreasing $\theta_\mathrm{res}$, that is, orbits that have left the vILR with a significant increase in $J_z$ (Fig.~\ref{fig:hist_trappead_resonance}).
    \item In terms of their dynamical properties (the value of the adiabatic invariant, the initial vertical action, and the frequencies), the orbits trapped in the vILR and passing through it are not practically different. However, they choose their own path, that is, they evolve differently.
\end{itemize}

\begin{acknowledgements}
We acknowledge financial support from the Russian Science Foundation, grant no. 24-22-00376. 
\par
We thank the anonymous referee for his/her thorough review and highly appreciate the very useful comments and suggestions that contributed to improving the quality of the article.
\par
We also acknowledge the use of the~\texttt{AGAMA}~\citep{agama} and ~\texttt{mpsplines}~\citep{Ruiz_Jose2022} \texttt{python} packages, without which the present work would not be possible.
\end{acknowledgements}
\bibliographystyle{aa}
\bibliography{main} 

\appendix

\section{Action-angle variables, the resonances}
\label{sec:appendix}

We estimate three \textcolor{black}{unperturbed} actions $\bm{J}=(J_\mathrm{R},\,J_z,\,L_z)$ and frequencies $\bm{\Omega} = (\kappa,\,\omega_z,\,\Omega)$ on different time scales using a subroutine similar to the one we used in our previous work \citep{Zozulia_etal2024}. 
First of all, the instantaneous actions and frequencies (calculated via {\tt AGAMA}'s  package subroutines, \citealt{agama}) experience periodic short-term fluctuations due to the influence of the bar on time scales of the order of one radial or vertical oscillation. 
We calculate the actions on these timescales  
by averaging them along the orbit from apocenter to apocenter ($J_\mathrm{R}$ and $L_z$) or from one $z$-maximum/minimum to another to obtain $J^\mathrm{max}_z$ and $J^\mathrm{min}_z$, respectively. The frequencies $\kappa$, $\omega^\mathrm{max}_z$, and $\omega^\mathrm{min}_z$ are defined as the reciprocal of the time between two adjacent apocentres ($T_\mathrm{R}$) or two $z$-maxima/minima ($T^\mathrm{max}_z$ and $T^\mathrm{min}_z$), multiplied by $2\pi$\footnote{Note that, unlike our previous work, we separately calculate the action and frequency along the $z$-axis in order to avoid highlighting any specific directions on the $z$-axis and correctly determine the initial phase of the angle $\theta_z$.}. Mean angular velocity $\Omega$ can be obtained through the increase in the polar angle calculated between two adjacent apocentres $\Delta \phi/T_\mathrm{R}$. After this, we have six piecewise functions (four for actions and four for frequencies) that can be smoothed using mean-preserving \textcolor{black}{spline} \citep{Ruiz_Jose2022}. This procedure allows one to obtain middle-term actions $\bm{J}(t)$ and frequencies $\bm{\Omega}(t)$ (see \citealp{Zozulia_etal2024} for details). Knowing $\bm{\Omega}(t)$, we also calculate the dependence of angles $\bm{\theta}=(\theta_\mathrm{R},\,\theta^\mathrm{max}_z,\,\theta^\mathrm{min}_z,\,\theta_\phi)$ on time, integrating expression for frequencies $\bm{\theta} = \int_{t_0}^t\bm{\Omega}(t)dt+\bm{\theta}_0$, where $\bm{\theta}_0$ is the angle phase at the time $t_0$. We use the fact that at the time of passage of the first apocenter  $\theta_R(t_0^\mathrm{apo})=0$ and $\theta_\phi(t_0^\mathrm{apo})=\phi_0$ ($\phi_0$ is the polar angle of the first apocentre). Also at the time of passage of the first $z$-maxima/minima $\theta^\mathrm{max}_z(t_0^{z_\mathrm{max}})=0$ and $\theta^\mathrm{min}_z(t_0^{z_\mathrm{min}})=\pm \pi$. It should be noted that the use of mean-preserving \textcolor{black}{spline} \citep{Ruiz_Jose2022} allows one to preserve the angle between two apocentres equal to $2\pi$ for $\theta_\mathrm{R}$, $\Delta \phi$ for $\theta_\phi$, and the angle between the two $z$-maxima/minima equal to $\theta_z$. We find the vertical variables as follows: $J_z = (J^\mathrm{max}_z+J^\mathrm{min}_z)/2$, $\omega_z = (\omega^\mathrm{max}_z + \omega^\mathrm{min}_z)/2$ and $\theta_z = (\theta^\mathrm{max}_z + \theta^\mathrm{min}_z)/2$. 
\par
In addition to the middle-term oscillations, there are also secular orbital changes. To study them, we average middle-term actions and frequencies over the corresponding oscillation period \citep{Zozulia_etal2024}. In this case, we repeated the averaging procedure, but it is carried out between the maximum and minimum values of all quantities ($\bm{\Omega}$ and $\bm{J}$).
\par
In this paper, we consider the adiabatic invariant of the following type $J_v=J_\mathrm{R}'+J_z+L_z/2$, where $J_\mathrm{R}'=J_\mathrm{R}-L_z \theta(-L_z)$ ($\theta(x)$ is the Heaviside step function). The introduction of $J_\mathrm{R}'$ variable is due to the fact, that the precession rate for flat ILR orbits with negative $\Omega$ ($\Delta \phi<0$) is equal to $\Omega_p=\Omega+\kappa/2$, and is equal to $\Omega-\kappa/2$ for orbits with positive one. With such a frequency modification, the adiabatic invariant for the flat orbits takes the form $J_f=J_\mathrm{R}-L_z/2$.

When an orbit is trapped into resonance, it starts to librate around some average angle \citep{Lichtenberg_Lieberman_1992, Binney_2020}, or near some stable point in the phase space. Let $\theta_{\mathrm{res}}$ be the resonant angle. Since we explore the relationship between vertical bar structure and vILR in this work, we define the resonant angle $\theta_\mathrm{res}$ as $\theta_z-\theta_\mathrm{R}$. Without loss of generality, we can select this angle in such a way that the stable point corresponds to $\theta_\mathrm{res}=0$ (or $\pi$), stable banana orbits (the orbit's tips are pointing downwards and upwards, respectively), or $\pi/2$ (or $3\pi/2$), unstable anti-banana orbits, \citep{Pfenniger_Friedli1991}. 
\par
Orbits can exhibit a rich variety of behavior. We were able to clearly identify four types of characteristic behavior. In practice, to identify resonant orbits, we assume that such an orbit can not leave the area between two adjacent stable angles, $0 \pm \pi/2$ ($\pi \pm \pi/2$) or $\pi/2 \pm \pi/2$ ($3\pi/2 \pm \pi/2$), respectively.
\begin{itemize}
    \item[1, 2.] The first two types of behavior are associated with a resonant angle circulation. If the resonant angle $\theta_\mathrm{res}$ consistently takes values that are multiples of $\pi/2$ three times in ascending or descending order, it corresponds to the circulation of the resonant angle with positive $\dot{\theta}_\mathrm{res}>0$ or negative $\dot{\theta}_\mathrm{res}<0$ angular velocities, respectively.
    \item[3.] The orbit librates near the stable point in the phase space. In this case, the orbit's resonant angle $\theta_\mathrm{res}$ crosses some angle multiple of $\pi/2$ at least 3 time without going beyond $\pm \pi/2$, we mark such an orbit as ``in resonance'' during the period of time between the first and last intersections of $0$ ($\pi$, $\pi/2$ or $3\pi/2$). In this case, the orbit makes at least one full revolution (libration) around a stable point on the phase plane.
    \item[4.] Passage through the resonance. In this case, the value of $\dot\theta_\mathrm{res}$ changes its sign.
    When the resonant angle $\theta_\mathrm{res}$ intersects $0$ ($\pi$, $\pi/2$ or $3\pi/2$) two times and then crosses $-\pi/2$ ($\pi/2$, $0$ or $\pi$) without completing a revolution on the phase plane, this means that the orbit changes its behavior and passes through resonance. We mark such an event as ``passage''. 
\end{itemize}

\par
Using our criteria for classifying orbital behavior, we not only determine whether the orbit is in resonance or in the passage, but we can also estimate the time it spends in one state or another (in terms of resonant angle changes).
\par
It should be noted that in this paper the notation  $\dot{\theta}_z \lessgtr \dot{\theta}_\mathrm{R}$ and $\omega_z/\kappa \lessgtr 1$ are not the same thing. The first inequality characterizes the behavior of only circulating orbits.
In the second case, $\omega_z$ and $\kappa$ are calculated numerically using the procedure described at the beginning of this section for all orbits, not just circulating ones.

\label{lastpage}
\end{document}